\documentclass{ws-ijmpa}
\usepackage[super,compress]{cite}
\usepackage{graphicx}

\def\Journal#1#2#3#4{{#1} {\bf #2}, #3 (#4)}
\def\PRL{\it Phys. Rev. Lett.}

\def\PRD{\it Phys. Rev. D}

\def\PLB{\it Phys. Lett. B}

\def\ChPC{\it Chin. Phys. C}
\def\SJNP{\it Sov. J. Nucl. Phys.}
\def\NPB{\it Nucl. Phys. B}

\begin{document}

\markboth{S.~Han, G.~Li, X.~Zhou, Q.~L.~Jie}{Initial state radiation correction and its effect to data-taking scheme for $\sigma^{\mathrm{B}}(e^+e^-\to ZH)$ measurement}

%
\catchline{}{}{}{}{}
%

\title{Initial state radiation correction and its effect to data-taking scheme\\for \boldmath$\sigma^{\mathrm{B}}(e^+e^-\to ZH)$ measurement}
 
\author{Shuang Han$^{1,\star}$, Gang Li$^{2,\dag}$, Xiang Zhou$^{1,\ddag}$, Quan-Lin Jie$^{1,\ast}$}

\address{$^1$ Hubei Nuclear Solid Physics Key Laboratory, Key Laboratory of Artificial Micro-\\ and Nano-structures of Ministry of Education, and School of Physics and Technology,\\ Wuhan University, Wuhan 430072, China\\
$^2$ Institute of High Energy Physics, Chinese Academy of Science, Beijing 100049, China\\
$^{\mathrm{\star}}$shuanghan@whu.edu.cn,\\ 
$^{\mathrm{\dag}}$li.gang@ihep.ac.cn,\\ 
$^{\mathrm{\ddag}}$xiangzhou@whu.edu.cn,\\
$^{\mathrm{\ast}}$qljie@whu.edu.cn}

\maketitle

\begin{history}
\received{25 Feb 2019}
\end{history}

\begin{abstract}
The measurement of Born cross section of $e^+e^-\to ZH$ process 
is one of the major goals of the future Circular Electron Positron 
Collider, which may reach a precision of 0.5\% at 240\,GeV. Such 
unprecedented precision must be guaranteed by both theoretical and 
experimental sides, such as the calculations of high order corrections, 
the knowledge of the $\sigma^{\mathrm{B}}(e^+e^-\to ZH)$ line shape. 
Uncertainty of the radiative correction factor at 240\,GeV 
caused by  the $\sigma^{\mathrm{B}}(e^+e^-\to ZH)$ line shape is 
evaluated in this work. Therefore, dedicated data-taking schemes 
are proposed in order to precisely calculate the ISR correction factor.

\end{abstract}

\ccode{PACS numbers: 12.15.-y, 13.66.Jn, 12.15.Lk}

\section{Introduction}

The historic observation of the Higgs boson in 2012 at the Large Hadron
Collider (LHC)~\cite{higgs_atlas,higgs_cms} declared the discovery
of the last missing piece of the most fundamental building blocks
in the Standard Model (SM). Although the SM has been remarkably successful
in describing experimental phenomena, a precision Higgs physics program
will be critically important given that the SM does not predict the parameters
in the Higgs potential, nor does it involves particle candidates for
dark matters. In particular, potential observable deviations of the
Higgs couplings from SM expectations would indicate new physics.
Therefore, the Higgs discovery marks the beginning of a new era of
theoretical and experimental explorations.

China has been investigating the feasibility of a high energy
Circular Electron Positron Collider (CEPC) as a Higgs factory
since 2013~\cite{CDR-A,CDR-D}. The CEPC will collide electrons 
and positrons at the center-of-mass energy of 240~GeV with an 
instantaneous luminosity of $3\times10^{34}$~cm$^{-2}$s$^{-1}$. 
With a clean environment, CPEC will provide a much clearer picture 
of the nature of Higgs and reveal many of the most profound 
mysteries intimately connected with the Higgs particle. The 
three leading Higgs production processes at a 240\,GeV CEPC 
are: Higgs-strahlung~($e^+e^-\rightarrow ZH$) and vector boson 
fusions~($e^+e^-\rightarrow\nu\bar{\nu}H$ and $e^+e^-\rightarrow e^+e^-H$). 
The CEPC is designed to collect 5.6~ab$^{-1}$ of integrated 
luminosity with two detectors in seven years, producing about 
$1.1\times10^6$ Higgs events.

One of the advantages at a $e^+e^-$ collider like CEPC is that 
the center-of-mass energy is precisely defined and many absolute 
measurements could be performed for Higgs boson. In a $ZH$ event, 
where the $Z$ boson decays to a pair of visible fermions~($Z\rightarrow 
e^+e^-,~\mu^+\mu^-,~\mbox{or}~q\bar{q}$), the Higgs boson can be 
identified with the kinematics of these fermion pairs independent 
its decays. It is claimed that the CEPC is able to measure the 
observed cross section ($\sigma^{\mathrm{obs}}$) of $e^+e^-\rightarrow ZH$ 
at 240\,GeV to a precision of 0.5\% by combining all three 
channels of $Z$ boson decays~\cite{CDR-D}. 

The Born cross section~($\sigma^\mathrm{B}$) at 240\,GeV, which is 
directly applicable to the theoretical analysis or independently 
comparing with results from other experiments, can then be obtained 
by applying corrections for initial state radiation (ISR) and 
other high order corrections. Unfortunately, the ISR correction 
at 240\,GeV depends on not only the theoretical calculations but 
also the line shape of the $e^+e^- \to ZH$ cross section, which 
needs to be constrained by experimental data. In this paper, the 
dedicated data-taking schemes for the radiative correction to 
the cross section of $e^+e^-\rightarrow ZH$ at 240\,GeV for CEPC 
is investigated. The data-taking schemes are optimized to collect 
data samples economically and effectively in order to achieve 
a significant better precision of the ISR correction factor at 
240\,GeV and to satisfy requirements experimentally and theoretically. 
Problems of determination of the center-of-mass energies and 
their integrated luminosities that need to be accumulated are 
carefully studied.

This paper is organized as follows: the ISR effect and theoretical
formulas for the radiative correction are described in Sec.~2, followed
by the procedure to calculate the radiative correction factor in
Sec.~3, the data-taking schemes are suggested in Sec.~4, and finally 
summary and discussion about the results are presented.

\section{ISR effect}

The ISR effect is an issue that cannot be avoided at $e^+e^-$ 
colliders. One of the incoming particles ($e^+$ or $e^-$) emits 
photon(s) before the interaction with the other, which reduces 
the beam energy prior to the momentum transfer. The ISR effect 
can be described with the structure function approach~\cite{isr-69,
isr-70,isr-71,Blumlein:2011mi,Blumlein:2019srk,isr-gm}, which yields 
an accuracy of 0.1\% due to the uncertainty of the radiative 
function $F(x,s)$. The uncertainty from $F(x,s)$ is neglected 
because it’s much smaller than the statistical uncertainty and 
could be further reduced with more theoretical work in the future. 
The experimental $\sigma^{\mathrm{obs}}$ of $e^+e^-$ colliders 
can be mathematically factorized as the integral of the Born 
cross section with the high order correction factors and $F(x,s)$,
\begin{eqnarray}\label{eqn-1}
\sigma^{\mathrm{obs}}(s)=\int^{1-s_{m}/s}_{0}\frac{\sigma^{\mathrm{B}}(s(1-x))}{|1-\Pi(s(1-x))|^2}F(x,s)~\mathrm{d}x,
\end{eqnarray}
\noindent where $\sigma^{\mathrm{B}}(s)$ is the Born cross
section at the center-of-mass energy $\sqrt{s}$ of the colliding
beam, and $\sqrt{s_{m}}$ in the upper limit of the integral
is the production threshold of the specific reaction,
and $1/|1-\Pi(s)|^2$ represents all the high order 
corrections~\cite{jiayu-1,lilinyang,jiayu-2}. Since 
the high order factors are independent of experiments 
and not issues to concern in this paper, it is dropped 
hereafter.

The ISR correction factor is defined to extract the Born
cross section from the observed one
\begin{eqnarray}\label{eqn-2}
1+\delta(s)=\frac{\sigma^{\mathrm{obs}}_{\mathrm{gen}}(s)}{\sigma^{\mathrm{B}}_{\mathrm{gen}}(s)}~,
\end{eqnarray}
\noindent where it should be noted that the 
$\sigma^{\mathrm{obs}}_{\mathrm{gen}}(s)$ and 
$\sigma^{\mathrm{B}}_{\mathrm{gen}}(s)$ are usually calculated 
with some dedicated generator(s) with some experiment-dependent 
kinematic cuts and measured $\sigma^{\mathrm{B}}$ line shape 
from the threshold up to $\sqrt{s}$ as inputs. Then the Born 
cross section at $\sqrt{s}$ can be determined by
\begin{eqnarray}\label{eqn-3}
\sigma^{\mathrm{B}}(s)=\frac{\sigma^{\mathrm{obs}}(s)}{1+\delta(s)}.
\end{eqnarray}

It should be noted that ISR correction factor is the 
function of center-of-mass energy, $\sqrt{s}$, and depends 
on not only theoretical calculations, but also experiment 
measurements. Furthermore its uncertainty directly contributes 
to the $\sigma^{\mathrm{B}}$. For the sake of convenience, 
$\Delta_{\mathrm{ISR}}$ is used to represent relative 
uncertainty of the ISR correction factor 
$[{\Delta(1+\delta)}/{(1+\delta)}]$ throughout the paper. 
In this paper, only $\Delta_{\mathrm{ISR}}$ at 240 GeV is studied, 
which is easy to  be replicated to other energy points.

\section{Calculation of ISR correction factor}

\subsection{Model independent measurement of $\sigma(ZH)$}

In the Higgs-strahlung process, the $e^+e^-$ annihilate
into a virtual $Z$ boson and becomes a real $Z$ by emitting a
Higgs boson, with the $Z$ boson mainly decaying
to a pair of fermions afterward. The center-of-mass energy is
precisely controllable at a $e^+e^-$ collider like the CEPC.
The Higgs boson can be identified with the recoil mass of
these fermion pairs with the following formula $m^{2}_\mathrm{recoil}
=(\sqrt{s}-E_{f\bar{f}})^2-p^2_{f\bar{f}}=s-2E_{f\bar{f}}\sqrt{s}
+m^{2}_{f\bar{f}}$, where $E_{f\bar{f}}$, $p_{f\bar{f}}$ and
$m_{f\bar{f}}$ are the energy, momentum, and invariant mass of
the fermion pair system, respectively. The $ZH$ event yield can be extracted
independently of the Higgs decays with the $m_\mathrm{recoil}$
spectrum.

Events with $Z$ decaying to $e^+e^-,~\mu^+\mu^-$, and $q\bar{q}$
are three ideal ways to identify the $e^+e^-\rightarrow ZH$
recoil mass spectrum and cover a majority of 76.6\% of the $Z$
decay modes. The observed cross section is calculated using 
\begin{eqnarray}\label{eqn-4}
\sigma^{\mathrm{obs}}(e^+e^-\rightarrow ZH)=\frac{N^{\mathrm{obs}}}{\mathcal{L}_{\mathrm{int}}\epsilon^{f\bar{f}}\mathcal{B}^{f\bar{f}}}~,
\end{eqnarray}
where $N^{\mathrm{obs}}$ is the total number of $ZH$ events 
observed, $\mathcal{L}_{\mathrm{int}}$ is the integrated luminosity 
accumulated at a certain $\sqrt{s}$, $\mathcal{B}^{f\bar{f}}$ 
is the branching fractions of $Z$ decaying to $e^+e^-,\,\mu^+\mu^-$, 
or $q\bar{q}$. The efficiency $\epsilon^{f\bar{f}}$ of event 
selection for the reaction is obtained by a full detector 
simulation and digitalization procedures~\cite{CDR-D}. 
The  major SM backgrounds are considered in the simulation and analysis 
as  references~\cite{CDR-D,zxchen} and more details 
on the Higgs signal and standard model backgrounds samples can be found in this paper~\cite{xinmo}. 
Then the three $Z$ decay modes are combined to form the final $\sigma^{\mathrm{obs}}(e^+e^-\rightarrow ZH)$ 
in order to improve the precision. 

\subsection{Method to extract ISR correction factor}
The ISR effect impacts on not only the production rate of $ZH$ process
but the shape of the recoil mass spectrum, which is used to determine 
the signal yield. Therefore, a full knowledge of ISR correction is 
essential for both measurements of $\sigma^{\mathrm{B}}(e^+e^-\to ZH)$ 
and Higgs boson mass. The expressions in Eq.~(\ref{eqn-1})-(\ref{eqn-3}) 
manifest mathematically that constraining the line shape of the Born cross 
section from production threshold to 240 GeV is needed to get a precise 
measurement of the ISR correction factor at 240\,GeV. The only feasible 
way is to collect a series of scan data samples between the threshold 
and 240\,GeV to constrain the line shape of 
$\sigma^{\mathrm{B}}(e^+e^-\rightarrow ZH)$.

\begin{figure}[b]
\centerline{\includegraphics[width=7cm]{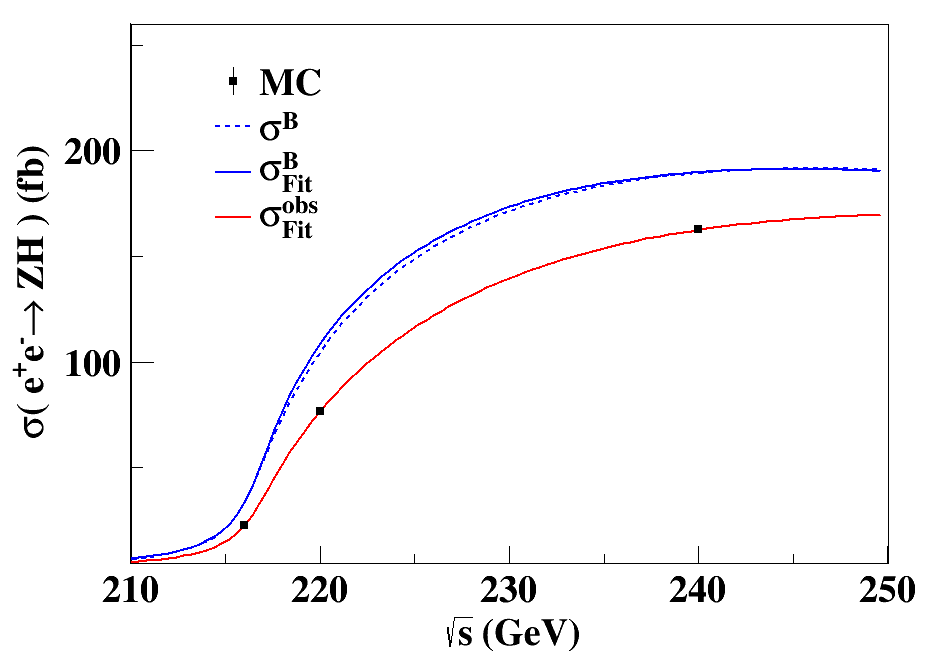}}
\caption{The fit to the MC samples generated at 216, 220, and 240\,GeV 
with luminosities of 0.2, 0.2, and 5.6~ab\,$^{-1}$, respectively. The 
blue dashed line refers to the $\sigma^{\mathrm{B}}(e^+e^-\rightarrow 
ZH)$ in the MC generator, the black foursquare markers with errors
refer to the observed cross sections calculated with MC samples, the 
blue and red lines are $\sigma^{\mathrm{B}}$ and $\sigma^{\mathrm{obs}}$ 
with the fitted parameters. \label{sm_fit}}
\end{figure}

The procedure described here is used to demonstrate the dependence 
of the ISR correction factor at center-of-mass energy of 240\,GeV 
on the line shape of $\sigma^{\mathrm{B}}(e^+e^-\to ZH)$. First, The Born cross 
section is assumed to be SM-like~\cite{xsection-uuh} and the radiative function, $F(x,s)$~\cite{isr-69,isr-70,isr-71,Blumlein:2011mi,Blumlein:2019srk}, is used to calculate the observed cross section. Then the MC signals 
of $e^+e^-\to ZH$ process are generated at 216, 220, and 240\,GeV with 
luminosities of 0.2, 0.2, and 5.6~ab\,$^{-1}$, respectively, and the 
background contributions are assumed to be the same as those at 240\,GeV 
because their cross sections change rather slowly in this energy region~\cite{xinmo}. Next, same analysis method~\cite{zxchen} is 
repeated to get observed cross sections and evaluate their statistical uncertainties. And next, the observed cross section of all energy points 
are fitted using Eq.~(\ref{eqn-1}) and $\sigma^{\mathrm{obs}}_{\mathrm{Fit}}$ and $\sigma^{\mathrm{B}}_{\mathrm{Fit}}$ 
are obtained simultaneously as shown in Fig.~\ref{sm_fit}. Finally, the 
ISR correction factor is calculated using Eq.~(\ref{eqn-2}):  
$(1+\delta)=\sigma_{\mathrm{Fit}}^{\mathrm{obs}}/\sigma^{\mathrm{B}}_{\mathrm{Fit}}$ 
at 240\,GeV, where $\sigma_{\mathrm{Fit}}^{\mathrm{obs}}$ and 
$\sigma^{\mathrm{B}}_{\mathrm{Fit}}$ are the best knowledge on 
the observed and Born cross sections and will be implemented into
generators.

\begin{figure}[b]
\centerline{\includegraphics[width=7cm]{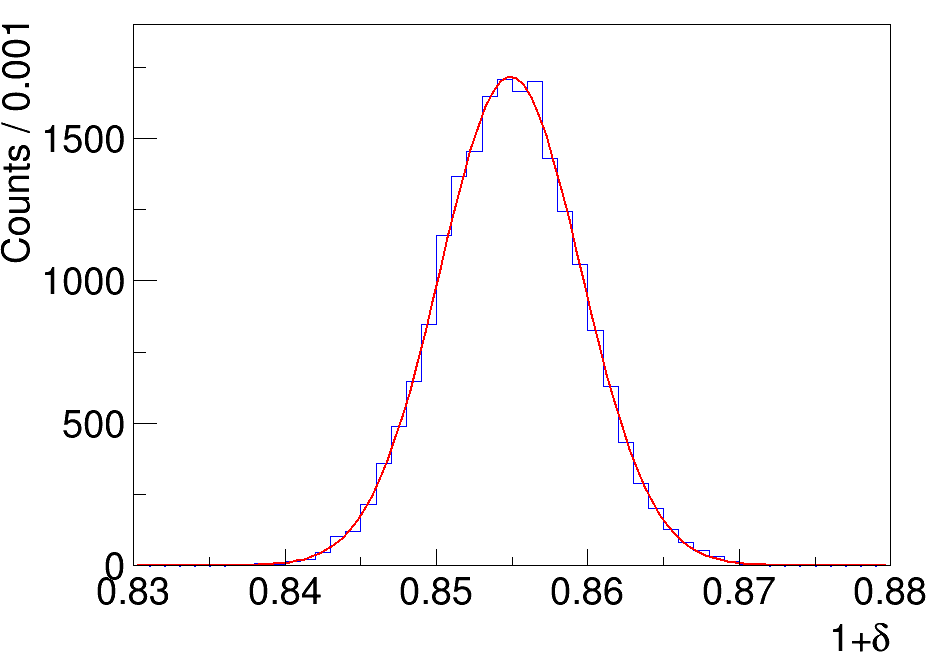}}
\caption{The fit result of the distribution of calculated $(1+\delta)$ 
using simulated samples at 216, 220, and 240\,GeV with luminosities of 
0.2, 0.2, and 5.6~ab$^{-1}$, respectively. The blue histogram is the 
distribution of 20,000 samplings. The red line is the fit result with 
a Gaussian function. The fitted mean value and standard deviation are 
0.855 and $4.6\times 10^{-3}$, respectively. \label{ISR_factor}}
\end{figure}

After repeat the above procedure 20,000 times, the distribution 
of the ISR correction factor at 240\,GeV is found to satisfy a Gaussian 
distribution as expected. The fitted mean and standard deviation 
are taken as the central value and uncertainty of the ISR correction 
factor, respectively. Fig.~\ref{ISR_factor} illustrates the fit results, 
and the relative uncertainty $\Delta_{\mathrm{ISR}}$ at 240\,GeV is 
0.54\% in this case.

\section{Optimization of Data-taking scheme}
In Eq.(\ref{eqn-1}), the integral kernel $\sigma^{\mathrm{B}}(s)$ 
is Born cross section, which should be measured by experiments.
The reason is that Higgs properties are still not yet solidly 
determined due to the limited experimental precision. On one 
hand, the present experimental results are sufficient to
discriminate between distinct hypotheses in Higgs boson spin
analysis~\cite{hspin-atlas,hspin-cms}. But on the other hand,
the determination of the $CP$ properties is in general much
more difficult, since in principle the observed state could
consist of any admixture of $CP$-even and $CP$-odd
components~\cite{hspin-cms,hcp-cms}. If physics is the SM, 
i.e., a single resonance with spin-0 and $CP$-even,
the Born cross section of the Higgs-strahlung process is
expected to reach its maximum at 250\,GeV approximately,
and then decreases with increasing center-of-mass energy.
From the experimental point of view, the center-of-mass
energies of the data samples determine the uncertainty
of the fitted line shape of the cross section. Besides,
the allocation of integrated luminosity, of the various
energy points could also make differences on the uncertainty 
of the ISR correction factor and of the $\sigma^{\mathrm{B}}$ 
at 240\,GeV. In conclusion, dedicated scan data samples 
between the threshold and 240\,GeV are needed to constrain 
the line shape of $e^+e^-\to ZH$ process.

For a measurement of 0.5\% statistical uncertainty, the 
sensitivity cannot be better than 0.5\% according to the 
definition in Eq.(\ref{eqn-1}) and (\ref{eqn-2}). In order to achieve 
a relative comparable precision of 0.5\% for the ISR correction 
factor at 240\,GeV, an economical and effective way of collecting 
data samples should be proposed for the CEPC. The effects 
of energies and allocation of integrated luminosity are 
investigated systematically in this section.

\begin{figure*}[htbp]
\centerline{\includegraphics[width=6.3cm]{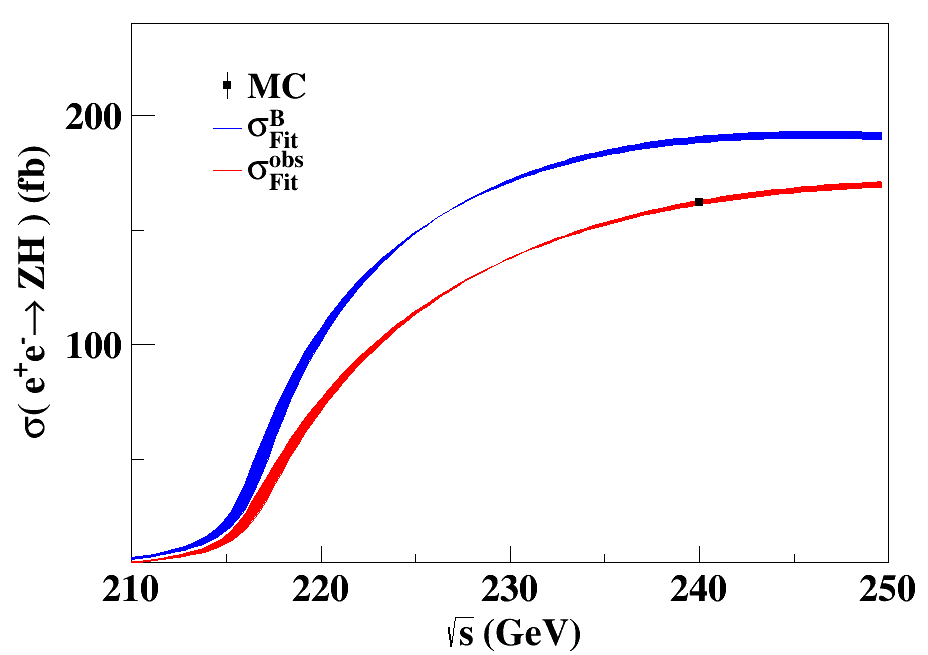}
\hspace{0.2cm}
\includegraphics[width=6.3cm]{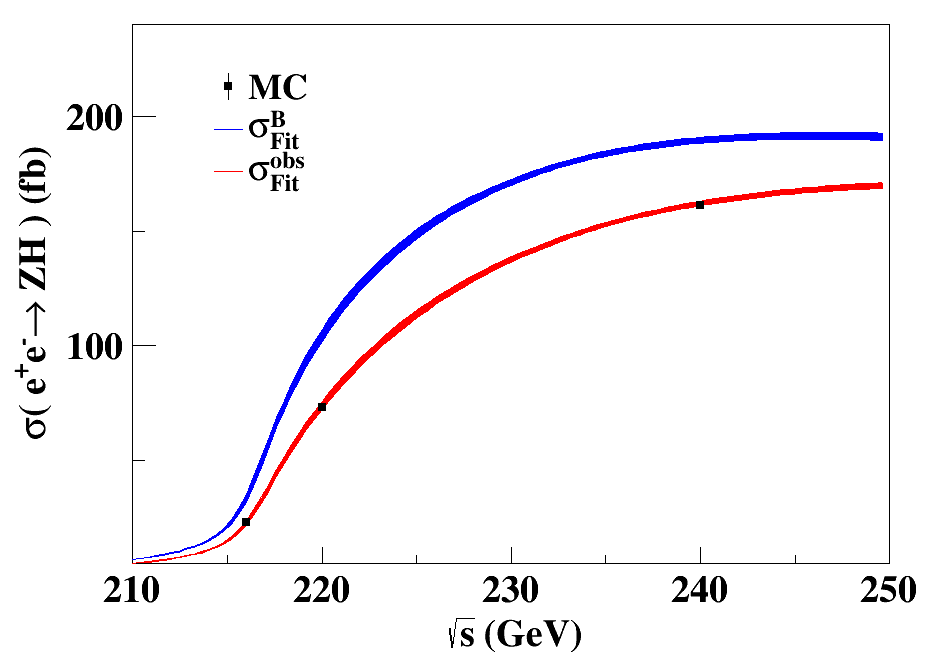}}
\caption{The fit results to MC samples, where the Born and observed 
cross sections are constrained to the blue and red bands. The left 
figure is based on one MC sample of 5.6~ab$^{-1}$ at 240\,GeV, and 
the right figure is based on three MC samples at 216, 220, and 240\,GeV 
with luminosities of 0.2, 0.2, and 5.6~ab$^{-1}$. \label{sm_fit_band}}
\end{figure*}

\subsection{Determination of the energies}
The impact of energies is studied by changing the combination of 
center-of-mass energies of MC samples. As a Higgs factory, CEPC is 
designed to accumulate a total of 5.6~ab$^{-1}$ integrated luminosity
running at 240\,GeV. The line shape of the Born cross section based 
on MC sample at this single energy point of 240\,GeV is shown in the 
left of Fig.~\ref{sm_fit_band}. The Born cross sections are constrained 
to the blue band with a bad performance in the low energy region which 
indicates that more data are in need below 230\,GeV. Our study shows 
that besides the established 5.6~ab$^{-1}$ data-taking plan at 240\,GeV, 
at least two more data samples at lower energy region are necessary 
to form a stable fit. With the luminosities fixed to 0.2\,ab$^{-1}$, 
MC samples are generated with center-of-mass energies varying from 
215 to 239\,GeV with a step of one GeV. Then fits are performed with 
the data sample at 240\,GeV together with MC samples randomly picked 
at two other lower energy points. As illustrated in the right of 
Fig.~\ref{sm_fit_band}, the Born cross sections are constrained 
to a much narrower band in the lower energy region.

\begin{figure}[b]
\centerline{\includegraphics[width=7cm]{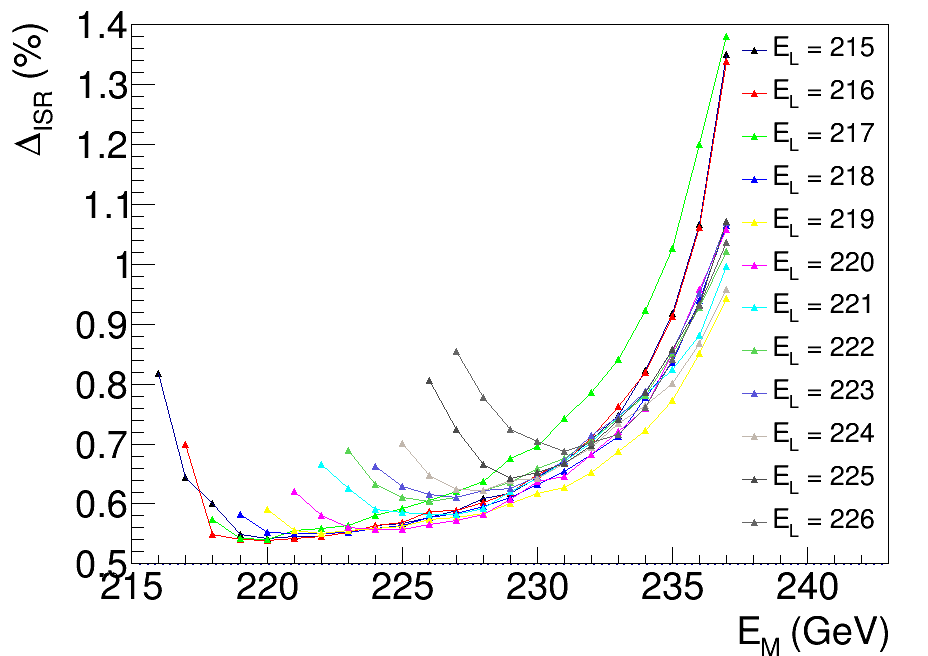}}
\caption{Fit results with three MC samples. 
Different colors denote fit results with MC samples at different 
lowest energies (in units of GeV) as shown in the legend. The 
horizontal axis is the center-of-mass energy for MC sample at 
middle energy point. The center-of-mass energy for the MC sample 
at the highest energy is fixed to 240~GeV. \label{energy}}
\end{figure}

\begin{table}
\tbl{Center-of-mass energies of three MC samples versus 
the uncertainty of the ISR factor at 240\,GeV. The luminosities 
for the three samples (energies from low to high) are fixed 
to 0.2, 0.2, and 5.6~ab$^{-1}$, respectively. \label{tab-1}}
{\begin{tabular*}{60mm}{@{\extracolsep{\fill}}cccc}
\toprule
\multicolumn{3}{c}{$\sqrt{s}$~(GeV)} & $\Delta_{\mathrm{ISR}}$~(\%) \\
\hline
215  &  220  &  240  &  0.54\\
216  &  220  &  240  &  0.54\\
217  &  220  &  240  &  0.54\\
218  &  222  &  240  &  0.55\\
219  &  222  &  240  &  0.55\\
220  &  224  &  240  &  0.56\\
221  &  227  &  240  &  0.58\\
222  &  226  &  240  &  0.61\\
223  &  227  &  240  &  0.61\\
224  &  228  &  240  &  0.62\\
\botrule
\end{tabular*}}
\end{table}

The most accurate fit results of the energy combinations with
the lowest energy of MC sample varies from 215 to 224~GeV are
listed in Table~\ref{tab-1} and the full-scale fit results can
be found in Fig.~\ref{energy}. As an example, the red points
refers to fit results with lowest energy MC sample generated
at 216~GeV, the horizontal axis is the center-of-mass energy
of MC sample at the middle energy point. Fit results with
different lowest energy MC sample are shown in the same figure.
The common feature is that $\Delta_{\mathrm{ISR}}$ decrease 
as a function of the center-of-mass energy of the MC sample 
at middle energy point, then increase above a certain point 
after three to four GeV interval approximately. The energy combination
is the most critical factor for the accuracy of $1+\delta$ at 240\,GeV.
For example, with the same luminosity combination of 0.2, 0.2,
and 5.6~ab$^{-1}$, the uncertainty can differ from 0.54\% to
14\%. Other luminosity allocations are studied, besides the
fit is also applied with more than three MC samples, we find
that three MC samples at 216, 220, and 240\,GeV gives a best
accuracy on average in case of the same integrated luminosity.

\subsection{Allocation of integrated luminosity}
\begin{figure}[b]
\centerline{\includegraphics[width=7cm]{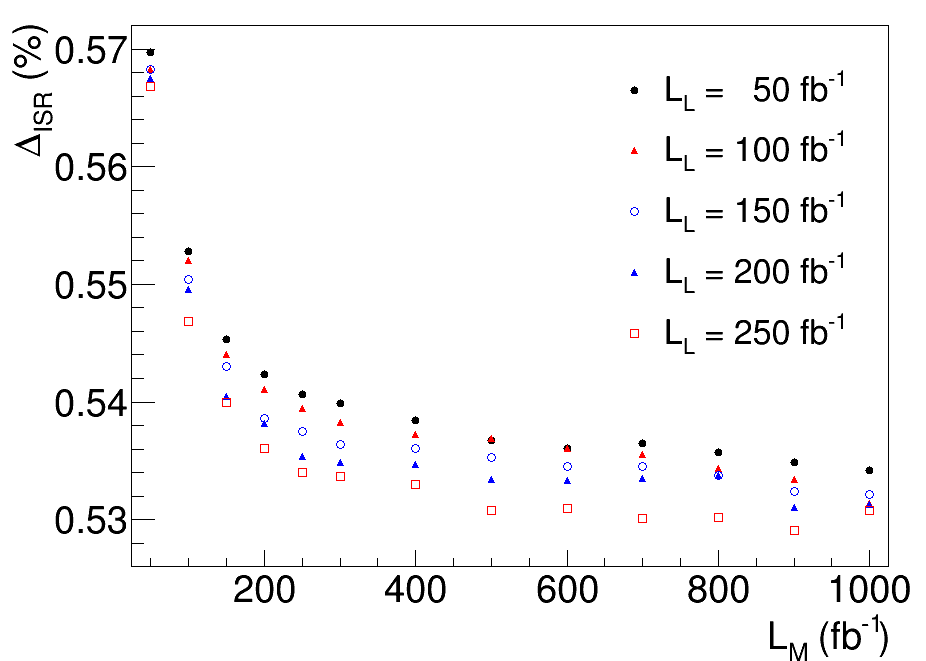}}
\caption{The fit results based on three MC samples
generated at 216, 220, and 240\,GeV. Different markers refer to 
MC samples at 216\,GeV of different luminosities as shown in the 
legend. The horizontal ordinate is the luminosity of MC sample at 
220\,GeV and luminosity for the MC sample at 240\,GeV is fixed to 
5.6~ab$^{-1}$. \label{lum}}
\end{figure}

The energies of three MC samples are fixed to 216, 220,
and 240\,GeV, which provides the best performance on average. 
With the luminosity for the MC sample at 240\,GeV fixed to
5.6~ab$^{-1}$, we change the luminosities of the other two MC
samples for the purpose of investigating the effect of statistics.
A significant improvement is that the relative uncertainty
$\Delta_{\mathrm{ISR}}$ are all below 0.6\% with the
energies fixed at 216, 220, and 240\,GeV as shown in Fig.~\ref{lum}.
The luminosity of the MC sample at lowest energy is fixed at 
a certain value, at first the uncertainty decrease rapidly as 
the luminosity of the MC sample at the middle energy increases. 
But the slope gradually reduce until finally reaching a plateau 
region where $\Delta_{\mathrm{ISR}}(240\,\mathrm{GeV})$ has a little reduction 
with the increase of the luminosity of the MC sample at the 
middle energy point. A comparison between results with different 
colors indicates that with a higher statistic for the MC sample 
at lowest energy, the slope of $\Delta_{\mathrm{ISR}}(240\,\mathrm{GeV})$
is bigger and gives a better averaged precision in the plateau
regions. However, in contrast with the significant improvement
due to a higher statistic of the MC sample at middle energy point,
the increase of the luminosity of the lowest energy MC sample does
not make significant changes.

\section{Summary and discussion}

In summary, the procedure of the calculation of ISR correction 
factor at 240\,GeV of the Higgs-strahlung process is investigated 
to match the statistical precision of the future CEPC. The effects 
of energies and statistics of data samples are studied systematically 
and the uncertainty of the ISR correction factor at 240\,GeV is 
evaluated accordingly. Based on the SM assumption, a economical 
and effective data-taking scheme is proposed. 

The study suggests an economical and effective proposal of
collecting data samples. With the established 5.6~ab$^{-1}$
data-taking plan at 240\,GeV as well as data samples at two
other energy points, the uncertainty for the ISR correction 
factor of $\sigma^{\mathrm{obs}}(e^+e^-\rightarrow ZH)$ at 
240\,GeV can be suppressed significantly. Selecting of energy 
points is the most critical factor for the precision, and 
it is found that three data samples at 216, 220, and 240\,GeV
give the best accuracy on average when fixing the total
integrated luminosity. The allocation of integrated luminosity 
can be found in Fig.~\ref{lum}. The study shows that higher 
priority should be given to 220\,GeV if the precision need 
to be improved further. The projected data sample (240\,GeV, 
5.6\,ab$^{-1}$) has dominant contribution to control the line 
shape of $\sigma^{\mathrm{B}}(e^+e^-\rightarrow ZH)$, because 
the lower energy regions count for a small fraction of the whole 
integral range in Eq.~(\ref{eqn-1}) and the Born cross sections 
are quite small comparing to higher energy regions. It should 
be noted that such scan data samples for the ISR correction 
are also useful to determine the Higgs boson spin and $CP$.

\section*{Acknowledgments}
This work was supported by the National Key Program 
for S\&T Research and Development (Grant No.: 2016YFA0400400), 
the Beijing Municipal Science \& Technology Commission 
project (Grant No.: Z1811000042180043) and the National 
Natural Science Foundation of China (Grant No.: 11205117, 
11575133).


\begin{thebibliography}{0}

\bibitem{higgs_atlas} ATLAS Collaboration (G.~Aad {\em et al.}), Observation of a new particle in the search for the Standard Model Higgs boson with the ATLAS detector at the LHC, \Journal\PLB{716}{1-29}{2012}.

\bibitem{higgs_cms} CMS Collaboration (S.~Chatrchyan {\em et al.}), Observation of a new boson at a mass of 125 GeV with the CMS experiment at the LHC, \Journal\PLB{716}{30-61}{2012}.

\bibitem{CDR-A} The CEPC Study Group, CEPC Conceptual Design Report: Volume 1 - Accelerator, {\ttfamily arXiv:1809.00285 [hep-ex]}.

\bibitem{CDR-D} The CEPC Study Group, CEPC Conceptual Design Report: Volume 2 - Physics \& Detector, {\ttfamily arXiv:1811.10545 [hep-ex]}.

\bibitem{isr-69} E.A.~Kuraev and V.S.~Fadin, \Journal\SJNP{41}{466-472}{1985}; Yad.~Fiz., \Journal\SJNP{41}{733-742}{1985}.

\bibitem{isr-70} G.~Altarelli and G.~Martinelli, CERN~\textbf{86-02},~47~(1986); O.~Nicrosini and L.~Trentadue, \Journal\PLB{196}{551-556}{1987}.

\bibitem{isr-71} F.A.~Berends, W.L.~Van~Neerven and G.J.H.~Burgers, \Journal\NPB{297}{429-478}{1988}.

\bibitem{Blumlein:2011mi} J.~Bl\"umlein, A.~De Freitas and W.~van Neerven, Two-loop QED operator matrix elements with massive external fermion lines, \Journal\NPB{855}{508-569}{2012}.

\bibitem{Blumlein:2019srk}
   J.~Bl\"umlein, A.~De~Freitas, C.~G.~Raab and K.~Sch\"onwald, The $O(\alpha^2)$ initial state QED corrections to $e^+e^-$ annihilation to a neutral vector boson revisited, \Journal\PLB{791}{206-209}{2019}.

\bibitem{isr-gm} M.~Greco, G.~Montagna, O.~Nicrosini, F.~Piccinini and G.~Volpi, ISR corrections to associated $HZ$ production at future Higgs factories, \Journal\PLB{777}{294-297}{2018}.

\bibitem{lilinyang} Y.Q.~Gong {\em et al.}, Mixed QCD-electroweak corrections for Higgs boson production at $e^+e^-$ colliders, \Journal\PRD{95}{093003}{2017}.

\bibitem{jiayu-1} Q.F.~Sun {\em et al.}, Mixed electroweak-QCD corrections to $e^+e^-\to HZ$ at Higgs factories, \Journal\PRD{96}{051301}{2017}.

\bibitem{jiayu-2} W.~Chen {\em et al.}, Mixed electroweak-QCD corrections to $e^+e^-\to\mu^+\mu^-H$ at CEPC with finite-width effect, \Journal\ChPC{43}{013108}{2019}.

\bibitem{zxchen} Z.X.~Chen {\em et al.}, Cross section and Higgs mass measurement with Higgsstrahlung at the CEPC, \Journal\ChPC{41}{023003}{2017}.

\bibitem{xinmo} X.~Mo {\em et al.}, Physics cross sections and event generation of $e^+e^-$ annihilations at the CEPC, \Journal\ChPC{40}{033001}{2016}. 

\bibitem{xsection-uuh} F.A.~Berends and R.~Kleiss, \Journal\NPB{260}{32-60}{1985}.

\bibitem{hspin-atlas} ATLAS Collaboration (G.~Aad {\em et al.}), Evidence for the spin-0 nature of the Higgs boson using ATLAS data, \Journal\PLB{726}{120-144}{2013}.

\bibitem{hspin-cms} CMS Collaboration (S.~Chatrchyan {\em et al.}), Study of the Mass and Spin-Parity of the Higgs Boson Candidate via Its Decays to Z Boson Pairs, \Journal\PRL{110}{081803}{2013}.

\bibitem{hcp-cms} CMS Collaboration (V.~Khachatryan {\em et al.}), Constraints on the spin-parity and anomalous $HVV$ couplings of the Higgs boson in proton collisions at 7 and 8\,TeV, \Journal\PRD{92}{012004}{2015}.

\end{thebibliography}
\end{document}